\documentclass[
 reprint,
 amsmath,amssymb,
 aps, physrev,
]{revtex4-2}

\usepackage{graphicx}
\usepackage{dcolumn}
\usepackage{bm}
\usepackage{lineno}

\usepackage[utf8]{inputenc}
\usepackage[T1]{fontenc}
\usepackage{mathptmx}
\usepackage{etoolbox}
\usepackage{xcolor}
\usepackage{hyperref}
\usepackage{float}
\usepackage[separate-uncertainty = true]{siunitx}
\usepackage{dcolumn}
\usepackage{bm}

\makeatletter
\def\@email#1#2{%
 \endgroup
 \patchcmd{\titleblock@produce}
  {\frontmatter@RRAPformat}
  {\frontmatter@RRAPformat{\produce@RRAP{*#1\href{mailto:#2}{#2}}}\frontmatter@RRAPformat}
  {}{}
}%
\makeatother
\begin{document}

\title[Spatial and energetic correlations of highly space-time-confined two electron pulses]{Spatial and energetic correlations of ultrashort two electron pulses}

\author{Stefan Meier}%
\thanks{These authors contributed equally.}
\affiliation{Department of Physics,  Friedrich-Alexander-Universität Erlangen-Nürnberg (FAU), 91058 Erlangen, Germany}

\email{stefan.m.meier@fau.de, peter.hommelhoff@fau.de}
\author{Jonas Heimerl}
\thanks{These authors contributed equally.}
\affiliation{Department of Physics,  Friedrich-Alexander-Universität Erlangen-Nürnberg (FAU), 91058 Erlangen, Germany}

\author{Felix López Hoffmann}%
\thanks{These authors contributed equally.}
\affiliation{Department of Physics,  Friedrich-Alexander-Universität Erlangen-Nürnberg (FAU), 91058 Erlangen, Germany}

\author{Tobias Volk}%
\affiliation{Department of Physics,  Friedrich-Alexander-Universität Erlangen-Nürnberg (FAU), 91058 Erlangen, Germany}

\author{Marco Knipfer}%
\affiliation{Department of Physics,  Friedrich-Alexander-Universität Erlangen-Nürnberg (FAU), 91058 Erlangen, Germany}

\author{Peter Hommelhoff}%
\affiliation{Department of Physics,  Friedrich-Alexander-Universität Erlangen-Nürnberg (FAU), 91058 Erlangen, Germany}
\affiliation{Faculty of Physics, Ludwig-Maximilians-Universität München (LMU), 80799 Munich, Germany}

\date{\today}
\begin{abstract}

When two electrons are emitted from a metallic needle tip into a nanometric volume on femtosecond timescales, strong Coulomb correlations arise. While longitudinal correlations, manifested as energy shifts, have been observed both from bare tips and in electron microscopes, transverse correlations remain hardly explored. Here, we present the first complete experimental characterization of such 3D correlations from needle tips. We find that electron pairs of small transversal spatial separation exhibit a pronounced energy gap of 3.3\,eV width, while electrons of small longitudinal energy separation show strong transverse repulsion, increasing their average mutual divergence angle by a substantial 34.5\%. The measurements are in excellent agreement with semiclassical point-particle simulations. These reveal that the maximal energy-gap magnitude is primarily determined by the laser pulse duration, whereas the maximal spatial separation is governed by the tip radius. The latter can be exploited in a remarkably simple method to generate strongly sub-Poissonian electron beams: A mere aperture, notably present in most electron-optical setups anyway, can act as an electron number sensitive filter. In combination with energy filtering we find an unprecedented suppression of multi-electron pulses reaching $g^{(2)}=0.02$ when keeping 3\% of the electron beam. These results provide a foundation for future studies of electron entanglement, correlated electron microscopy, and the design of ultrafast electron-optical instruments.

\end{abstract}
\maketitle

\section*{Introduction}
Correlations between elementary particles are a cornerstone of modern quantum mechanics. Since the introduction of higher order correlation functions by Glauber \cite{Glauber2006a}, they have formed modern quantum optics, probing quantum properties not only of photons, but also of electrons in two-dimensional electron gases \cite{Henny1999}, atoms \cite{Schellekens2005}, and those of elementary particles \cite{baym1998physics,Iannuzzi2006}. These correlation functions offer a direct way to study fundamental particle interactions. For electrons, the second-order correlation function can show anti-bunching both due to the fermionic nature of the electrons and Coulomb repulsion. These effects were studied with continuous free electron beams \cite{Kiesel2002,Kodama2011}, and nanosecond pulsed ones \cite{Kuwahara2021}. Recent studies have focused on investigating and harnessing electron-electron interactions of ultrashort electron pulses for the generation of sub-Poissonian electron beams \cite{Keramati2021,Meier2023,Haindl2023, Borrelli2024,Kuttruff2024}. These free electron beams are typically generated from nanometric electron sources, such as metal needle tips or Schottky emitters, by liberating electrons either by field electron emission or by (ultrafast) laser-triggering. When using femtosecond laser sources in particular, the electron-electron interaction after emission is so strong that clear anti-correlation signatures in energy \cite{Meier2023,Haindl2023} or in time \cite{Kuttruff2024} appear.
Theoretical investigations suggest that these interactions might even induce entanglement between two electrons \cite{Schattschneider2020}. This entanglement is currently investigated in transmission electron microscopes by introducing a second interaction zone for the electron pair \cite{Haindl2025,Tziperman_2026}.
However, one issue these experiments in electron microscopes face is the strong spatial filtering that is imposed between the emitted electrons and the detected electrons.
The interaction between the detected electrons and the many more emitted electrons will likely affect the quantum state of the electrons. For this reason, we study the emission of the electrons without any spatial filter between the detector and the bare tip.
We use a time-of-flight spectrometer equipped with a delay-line detector that offers the possibility of investigating spatial and temporal, i.e., energetic, correlations simultaneously \cite{Meier2022,Knipfer2024}. 

\begin{figure}
    \centering
    \includegraphics[width=\linewidth]{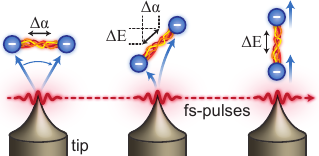}
    \caption{\textbf{Fully resolved two-electron correlation:} Multi-electron events are emitted after optical excitation with an ultrashort laser pulse from a metal needle tip. If more than one electron is emitted, they interact via Coulomb interaction, leading to a 3D repulsion (center). Measuring the full momentum of two electrons simultaneously, a competition between transverse repulsion, which increases their mutual angle $\Delta \alpha$ (left), and longitudinal repulsion, increasing their energy difference $\Delta E$ (right), can be observed. The exact 3D shape of this correlation map is measured by a delay-lined-detector.}
    \label{fig:figure1_setup}
\end{figure}

\begin{figure*}
    \centering
    \includegraphics[width=1\linewidth]{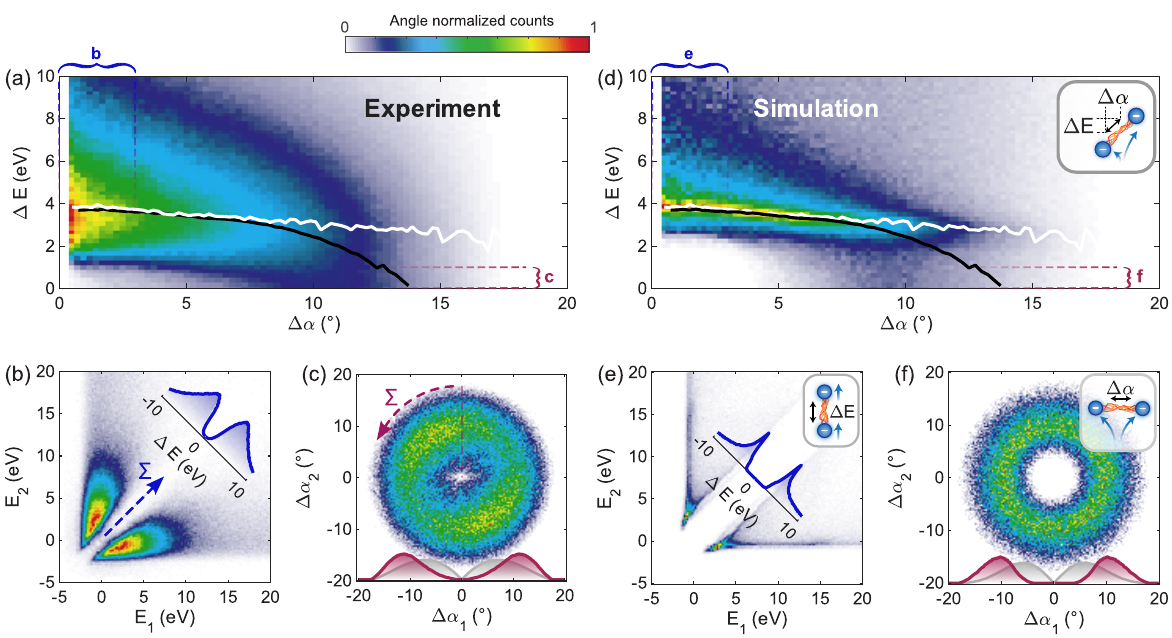}
    \caption{\textbf{3D-resolved electron correlations in experiment and simulation}. (a) Angular energy-difference spectrum (AEDS) map recorded with laser pulses with a pulse duration of \SI{12}{\femto\second}. The geometric suppression of counts towards small angles is counteracted by dividing column-wise by $\Delta \alpha$. The energy-repulsion-line (ERL, column-wise maximum line) is depicted for both the experimental distribution (black line) and the simulation (white line). 
    (b) Angular-filtered distribution [\,$\Delta \alpha<\SI{3}{\degree}$, region in (a) labeled (b)] shown as map with the energy of one electron depicted on the horizontal axis and the energy of the other on the vertical axis (static bias voltage ($e\cdot V$) contribution is subtracted). Because of the chosen angular filter, this configuration dominantly results in longitudinal repulsion. The inset shows the diagonal marginal, i.e., the energy difference distribution, with a pronounced energy gap with a width of \SI{3.3}{\electronvolt} (half peak-to-peak distance).
    (c) Angular map of the energy-filtered distribution [$\Delta E < \SI{1}{\electronvolt}$, i.e., region in (a) labeled (c)]. A substantial suppression of two-electron events for small angle differences is clearly visible. The red curve at the bottom shows the radial marginal distribution. For comparison, the corresponding distribution obtained from synthetically paired, and therefore uncorrelated, electrons is shown in gray.
    (d) Simulated AEDS-map. Both simulated ERL (white line) and experimental ERL (black) are overlayed. (e) Energy-correlation map as in (b); the thin horizontal and vertical features not showing up in the experimental spectra result from a simulation artifact. (f) Angular correlation map as in (c). Most features match well both qualitatively and quantitatively.}
    \label{fig:figure2}
\end{figure*}

We show experimental results of Coulomb-correlated electrons photoemitted from nanometer sharp metal needle tips, in particular. We present a fully three-dimensional analysis of the Coulomb interaction, demonstrating that a confined emission in time and space leads to either a strong energetic or a strong spatial repulsion, with a smooth transition between both cases (Fig.\ref{fig:figure1_setup}). With the help of semi-classical simulations, we show that the precise spatio-temporal confinement of the emission process is crucial for the electron-electron correlations. We show that not only energy but also spatial filtering allows us to shape the electron number distributions of free electron beams, and that combining both approaches the tunability is strongly enhanced.

\section*{Experimental results on 3D resolved electron correlations}
In the experiment, we trigger electrons from metal needle tips using laser pulses from an amplified erbium fiber oscillator. After pulse picking at \SI{100}{\kilo\hertz} and pulse compression with a highly non-linear normal dispersion fiber \cite{Lesko2021}, light pulses with a pulse duration of \SI{12}{\femto\second} at a central wavelength of \SI{1570}{\nano\meter} are available. The pulses are focused by an off-axis parabolic mirror with a focal length of \SI{15}{\milli\meter} to a spot size of \SI{3}{\micro\meter} inside an ultra-high vacuum chamber (base pressure of \(\SI{8e-10}{\milli\bar}\)), where the metal needle tip is situated. The focus position is aligned to the tip apex, which has a radius of curvature of approximately \SI{10}{\nano\meter}. After the electron emission in a non-linear photoemission process, the electrons are accelerated away from the tip with the help of a small bias voltage of around \SI{-20}{\volt} applied to the tip. The electrons propagate freely towards a delay-line detector (DLD), which can measure up to four electrons simultaneously, registering the time-of-flight and the position of each individual electron. Based on a neural network, we have improved the multi-hit capability of the detector such that the dead-radius of the detector shrinks to less than \SI{2}{\milli\meter} for events where two electrons are detected simultaneously \cite{Knipfer2024}. In this section, we investigate only two-electron events. We analyze the electron correlations in full 3D by analyzing the momentum differences of the electrons \(\Delta \vec{p}=\vec{p}_2-\vec{p}_1\). For convenience and comparability to other (electron microscopy) experiments, we calculate angular energy-difference spectra (AEDS) given by the energy difference \(\Delta E\) plotted versus the angle $\Delta\alpha$ between the final 3D momentum vectors of the two electrons.\\
In the experiment, we use a near-field intensity of $\sim$\SI{1.7e13}{\watt\per\square\centi\meter}, corresponding to a ponderomotive energy of $U_\mathrm{P} = \SI{4}{\electronvolt}$ and a Keldysh parameter of roughly $\gamma = 0.7$. These parameters indicate that the electrons are emitted on a sub-cycle time scale and are strongly driven by the optical field after the emission. We obtain a total event rate of $\sim$\SI{28}{\kilo\hertz}, of which $\sim$\SI{17}{\percent} are two-electron events.\\
In Fig.~\ref{fig:figure2}(a), we show an AEDS map. For better visibility, the uncorrelated, purely geometric decrease of counts at small angle differences is counter-acted by normalization to differential angular area, i.e., by column-wise division by the $\Delta \alpha$. In addition, we highlight the maximum value for each $\Delta \alpha$, resulting in an energy-repulsion line (ERL, black line), which is not affected by the described angle-normalization. For small values of \(\Delta\alpha\) the ERL indicates the maximum gap size of \(\approx\SI{3.3}{\electronvolt}\). This repulsion decreases as $\Delta \alpha$ increases, as shown by the ERL.\\
Two limiting cases in this map are particularly important, as they can be realized by filtering the electron beam in energy or location (e.g., through an aperture). We show the case of spatial filtering ($\Delta\alpha<\SI{3}{\degree}\)) in Fig.~\ref{fig:figure2}(b) and the case of energy filtering (\(\Delta E<\SI{1}{\electronvolt}\)) in Fig.~\ref{fig:figure2}(c).\\
We first demonstrate angular filtering and the resulting energy correlations. We select electrons with an angular difference of $\Delta \alpha < 3^\circ$ (Note that values below 0.5° are neglected due to artifacts within the detector dead radius, but anyway barely contribute due to geometric suppression \cite{Knipfer2024}). The spectral map of the resulting two-electron events are shown in Fig.~\ref{fig:figure2}(b), where the horizontal axis is the energy of one electron and the vertical of the other. We observe a strong suppression of two-electron events along the diagonal in the energy correlation map, similar to previous studies \cite{Meier2023,Haindl2023,Kuttruff2024}. Yet in contrast to these works, the energy distribution extends over a broad energy range beyond the range of the plot up to  $\sim$\,\SI{40}{\electronvolt} because of the energy plateau of the rescattered electrons~\cite{Meier2023}.\\
For the other limiting case, we filter electrons with an energy difference of $\Delta E < \SI{1}{\electronvolt}$ (horizontal box in (a)). In Fig.~\ref{fig:figure2}(c) we show the angle difference $\Delta\alpha_1$  along one axis and $\Delta \alpha_2$ along the other axis for these energy-filtered electron-pairs. We observe doughnut-shaped pattern with close to zero count rate in the center, demonstrating a clear suppression of two electrons that are close to each other spatially. The radial marginal distribution (red line) shows that the probability of detecting two electrons reaches a maximum at \(\Delta \alpha = \sqrt{\Delta\alpha_1^2+\Delta\alpha_2^2}=\SI{10.9}{\degree}\)and exhibits a half width at half maximum (HWHM) of approximately \SI{4.5}{\degree}. For comparison, an analogous distribution can be constructed from synthetically paired, and therefore uncorrelated, single-electron events (gray line). The comparison reveals the enhanced divergence of correlated electron pairs, whose average mutual divergence angle is 34.5\% larger than that of the uncorrelated reference distribution.\\
To deepen our understanding of these experimentally observed two-electron correlations, we carried out detailed numerical simulations, based on a point-particle trajectory model (for details see~\cite{Meier2023,Meier2024}). In the simulation, we model the tip surface as a hemisphere and vary the initial conditions of each emitted electron such that a similar emission probability is achieved. The number of emitted electrons per pulse follows a Poissonian distribution. After an electron is emitted, we let it interact with the other emitted electrons.\\
In Fig.~\ref{fig:figure2}(d) we show the so-obtained simulated AEDS map and the simulated ERL (white line) together with the experimental one (black). We chose parameters matching the experiment, in particular a central wavelength of \(\lambda=\SI{1550}{\nano\meter}\), a near-field intensity of \SI{2.2e13}{\watt\per\square\centi\meter}, a pulse duration of \SI{12.5}{\femto\second} and a tip radius of \SI{15}{\nano\meter}. Similar to the experiment, we observe that the ERL decays for increasing $\Delta \alpha$. For $\Delta \alpha = 0$, we observe a maximum energy gap size of \SI{4}{\electronvolt}, fairly well matching the experimental \SI{3.3}{\electronvolt}.\\
In Fig.~\ref{fig:figure2}(e,f), we show the same limiting cases 
as for the experiment (b,c). In the simulation, the probability to detect two electrons is maximal at \(\Delta \alpha = \sqrt{\Delta\alpha_1^2+\Delta\alpha_2^2}=\SI{9.7}{\degree}\) with a width of \SI{3.0}{\degree} (HWHM). Both maps show the same core feature as the experiment: the strong suppression of two-electron events along the diagonal in energy and the doughnut-like structure in the angle differences.
However, from the comparison of experiment (a-c) and simulation (d-f), we see that the simulated results (d-f) appear sharper and less smeared-out as well as the contrast of the repulsion is even more pronounced. This is likely because the electrons in the experiment are smeared out due to their wave packet nature, together with the finite resolution of the detector.\\
The high level of agreement between experiment and semi-classical simulation suggests that correlation effects arising from the fermionic nature of electrons (Pauli blocking)~\cite{Lougovski2011} or Coulomb blockade phenomena in and outside the metal play only a minor role in the multi-electron emission process under the conditions considered here. Instead, the overall shape of the anti-correlation signal is governed primarily by classical point-particle dynamics rather than wave-packet-related effects. In other words, although the emission process of each individual electron is inherently quantum-mechanical, the mutual dynamics outside the tip appear to be dominated by the motion of their respective centers of mass, illustrating the quantum-to-classical cross-over via the Ehrenfest theorem, both in the longitudinal and the transverse direction.\\

\section*{Parameter scalings}
\begin{figure*}
    \centering
    \includegraphics[width=0.99\linewidth]{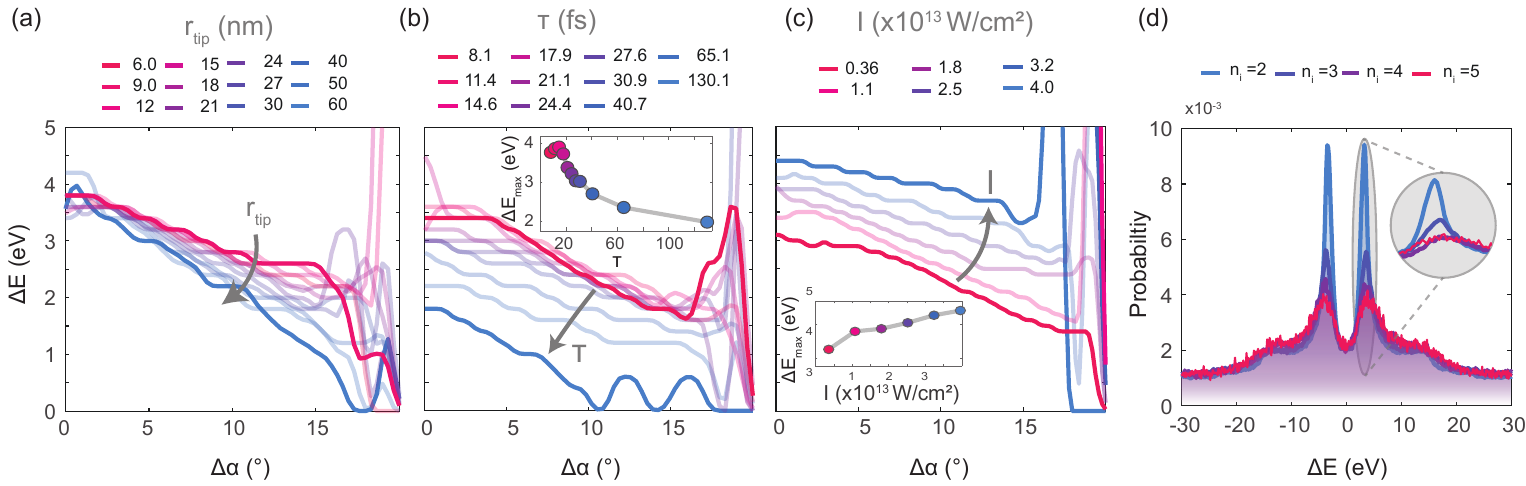}
    \caption{\textbf{Parameter dependencies of the energy repulsion line (ERL)}. Simulated ERLs for a near-field intensity of \SI{2.2e13}{\watt\per\square\centi\meter}, a pulse duration of \SI{12.5}{\femto\second} and a tip radius of \SI{15}{\nano\meter}, where one parameter is then varied. (a) When changing the tip radius $r_{\mathrm{tip}}$ from \SI{6}{\nano\meter} to \SI{60}{\nano\meter}, the maximum gap size stays almost constant and mainly the slope of the ERL changes. (b) For an increasing pulse duration $\tau$ from \(\sim\SI{8}{\femto\second}\) (red) to \(\sim\SI{130}{\femto\second}\) (blue), the maximum gap size $\Delta E_{\mathrm{max}}$ overall decreases (see inset), while the slope stays similar. In the limit of below two optical cycles of pulse duration, the maximum gap size is almost constant.  (c) An increasing intensity $I$ from $\SI{0.36e13}{\watt\per\square\centi\meter}$ to $\SI{4.0e13}{\watt\per\square\centi\meter}$ leads to both a larger gap size and a flatter ERL. The maximum gap size increases from \SI{3.3}{\electronvolt} at an intensity of \SI{0.36e13}{\watt\per\square\centi\meter} to \SI{4.4}{\electronvolt} at \SI{4.0e13}{\watt\per\square\centi\meter} (see inset). In (a-c), the large fluctuations for $\Delta \alpha \geq 15°$ are artifacts due a small count rate for large angle differences because of the half-opening angle of the emission profile. (d) Energy difference histograms of events where two electrons are registered at the virtual screen but $n_i=2$ (blue) to $n_i=5$ (red) electrons were initially emitted at the tip. The coarse-shape of the Coulomb-induced gap remains for all cases similar, but the contrast of the gap decreases for events that initially started with more electrons (see text for details).}
    \label{fig:figure3}
\end{figure*}

Resting on the good match between experiment and  simulation, we can now study how the correlation effects behave when we vary important experimental parameters, like the tip size, the pulse duration and the intensity, and how undetected electrons affect the results.\\
When we change the tip radius in the simulation, we find that the maximum energy gap stays almost constant for small angles, but the decay of the ERL towards larger \(\Delta\alpha\) is steeper for larger tips (Fig.~\ref{fig:figure3}(a)). We note that the size of the emission area at the tip (the part of the tip from which the electrons are emitted) does not influence the shape of the ERL, but only redistributes the local intensity in the AEDS map. From an intuitive point, we can explain the maximum gap size by events that fly almost on the same line to the detector, hence, repel each other in energy (longitudinally) mainly. This situation is realized in electron microscopy setups due to pronounced on-axis spatial filtering, which explains why similar results are obtained for a comparable pulse duration~\cite{Haindl2023,Tziperman_2026}.

When we change the pulse duration (Fig.~\ref{fig:figure3}(b)) for a fixed tip radius of \SI{25}{\nano\meter}, an increasing pulse duration leads to an overall reduction of the maximum energy gap size from around \SI{3.6}{\electronvolt} for $\tau = \SI{8.1}{\femto\second}$ to $\sim$\SI{1.8}{\electronvolt} for $\tau = \SI{130}{\femto\second}$ (see also inset in (b)). This reduction is similar for all \(\Delta \alpha\), i.e., it looks like a parallel shift of the ERL to smaller values of \(\Delta E\). This result agrees with previous observations that a stronger confinement in time generally leads to a stronger energy repulsion~\cite{Meier2023,Haindl2023}. For extreme pulse durations like $\tau = \SI{8.1}{\femto\second}$, i.e., below two-optical-cycle pulses, this trend does not continue. In this extreme case, it is likely that both electrons are emitted in only one of the cycles and thus lead to dominantly spatial correlations. The maximum energy gap hence seems to be limited. Quantum-mechanical simulations are required for these cases \cite{Classen2025}, which are beyond of the scope of this work.

Our simulation shows that a similar impact on the shape of the ERL can be achieved by increasing the laser intensity. Here we observe that the maximum gap size increases from $\SI{3.1}{\electronvolt}$ to $\SI{4.4}{\electronvolt}$ by increasing the intensity by a factor of ten (see inset of (c)). Simultaneously, the slope of the ERL becomes slightly flatter. This is caused by both the change of the emission rate from an intensity envelope dominated to a sub-cycle time scale emission rate, together with the increased driving of the electrons for higher intensities.

Lastly, we can study the influence of undetected electrons based on our simulation. The number of emitted electrons initially follows a Poissonian distribution. After propagation we then exclude all electrons that did not hit the (virtual) detector in the simulation to mimic the real experiment. In Fig.~\ref{fig:figure3}(d) we show the energy difference histogram for all simulated two-electron events hitting the screen, labeled by the number of initially emitted electrons within the respective emission event. The overall shape and gap size is barely dependent on the number of initially emitted electrons, but the distribution becomes smeared out subsequently. The effect is the same if we introduce an artificial finite detection efficiency, i.e., when we randomly delete a certain fraction of the electrons reaching the screen. We can hence argue that the main interaction is always a two-electron interaction with the possible additional electrons acting as disturbance that smear out the measured distributions.

We note that, in general, achieving complete quantitative agreement between experiment and simulation is challenging and not always unique, as many parameters influence the shape of the anti-correlation signal. 
Using the ERL as a precise measure for the emission rate or the tip radius can hence be misleading. However, our simulations show how one can shape the ERL such that a maximum or minimum gap size can be achieved. This knowledge is especially important when one is interested in controlling the reduction of Coulomb effects to achieve small aberrations, for example, or to enhance them for shot noise-reduced electron beam generation and imaging, as we show next.

\section*{Energy and Spatial filtering}
\begin{figure*}
    \centering
    \includegraphics[width=0.99\linewidth]{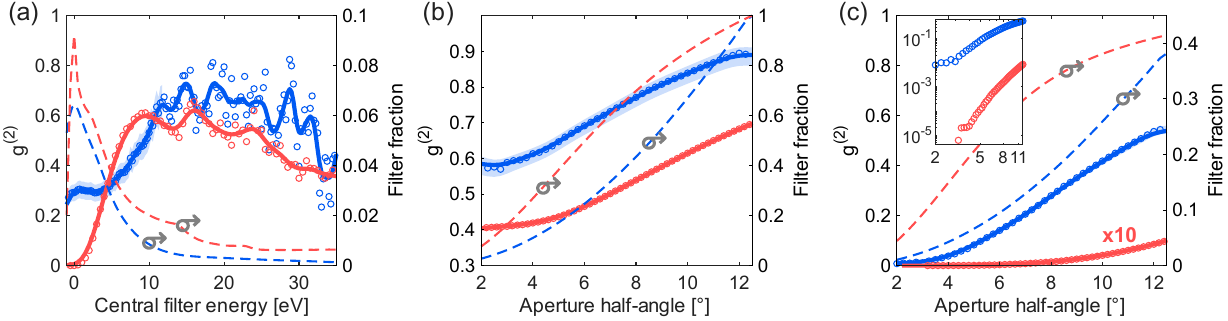}
    \caption{\fontsize{11}{12}\selectfont{\textbf{Second-order correlation function of experimental data and simulation after post-processing filter}. (a) $g^{(2)}$ as function of the filter central energy with a window size of \SI{1}{\electronvolt} for experiment (blue circles) and simulation (red circles). Solid lines show a Gaussian-weighted moving average with $\sigma=0.5$\,eV. The right-hand axis shows the fraction of electrons in that energy filter bin for the experiment (blue dashed) and the simulation (red dashed), normalized to the rate impinging the full detector surface. Conservative boundaries on the detector's losses allow to provide an error bound for $g^{(2)}$ (shaded area) for low energies, while for high energies, event rates become low and the $g^{(2)}$ estimate is shot-noise dominated. (b) Analogous depiction of $g^{(2)}$ as a function of a virtual aperture-opening half-angle, for a second measurement. (c) Dataset of (b) applying an additional energy filter window of 2\,eV width, centered at 1\,eV. The $g^{(2)}$ values of the simulated filtered beam are scaled for better visibility. The logarithmic depiction in the inset illustrates the onset of noise in the $g^{(2)}$ estimate due to low count numbers at small filter angles.}}
    \label{fig:figure4}
\end{figure*}

It is well known that the electron repulsion and the ensuing energy gap allows us to change the electron number statistics of the electron beam from Poissonian to sub-Poissonian statistics by simple energy filtering \cite{Meier2023,Haindl2023}. We repeat this filtering approach, but now deep in the strong-field regime and thus over a large energy range of more than 30\,eV. When we post-select the electron events with an energy pass filter of width $\SI{1}{\electronvolt}$, we observe a second-order correlation function as low as $g^{(2)} = 0.29$ for zero energy offset, i.e., at the energy given by the bias voltage (see Fig.~\ref{fig:figure4}(a), blue line), corresponding to the directly emitted electrons. For higher central filter energies, $g^{(2)}$ rises and reaches a broad plateau over the entire spectral region dominated by rescattered electrons. The blue shaded area corresponds to an error boundary derived from conservative estimates of various losses in the detector (see appendix \ref{sec:lossAppendix}). In the plateau region, the number of 2 electron events with both electrons within a 1\,eV window become so few, that shot noise dominates the uncertainty in the $g^{(2)}$ calculation, hence inhibiting the error bound calculation. Importantly the resulting noise on the $g^{(2)}$ curve is low enough to suggest that this plateau really lies systematically below 1. The overall behavior of $g^{(2)}$ is surprisingly well matched by the simulations (red line). The observations indicate a strong interaction of direct electrons and a reduced impact of Coulomb interaction due to the strong driving of the optical light field on rescattered electrons. The dashed lines (right axis) indicate the fraction of beam electrons transmitted through each 1\,eV-wide energy filter window.

In a similar manner to the energy filter, an aperture can be used to generate sub-Poissonian electron beams, an effect which notably will arise in an electron microscope naturally \cite{Haindl2023}. The aperture limits the maximum angle difference and the angle relative to the electron-optical axis (aperture half-angle) at the same time. In order to cover a larger range of angles we performed a second measurement with a smaller tip-to-detector distance (see appendix \ref{sec:meas2Appendix}).We simulate the effect of a varying aperture again by post-selecting the electrons. The resulting $g^{(2)}$ behavior is shown in Fig.~\ref{fig:figure4}(b). Applying a post-selection aperture half-angle of \SI{2}{\degree} results in a sub-Poissonian electron beam with $g^{(2)} = 0.57$ in the experiment (blue solid line) and $g^{(2)} = 0.40$ in the simulation (red solid line). For larger angles the second-order correlation function slowly approaches a Poissonian distribution, and would reach $g^{(2)} = 1$ for an infinite aperture, as all electrons are collected again. In the experiment an unknown fraction of the beam is lost outside the detector surface, so the fraction of electrons passing the post-selection filter (right axis dashed lines, blue experiment, red simulation) are normalized to the maximum number of electrons within the detector area. The dark shaded area corresponds to error boundaries estimated similarly as in Fig.~\ref{fig:figure4}(a). Crucially, the average number of post-selected electrons per pulse is much larger than in the narrow energy windows above, so the contribution of three-electron events to the value of $g^{(2)}$ is not negligible anymore. These boundaries demonstrate that the decrease of $g^{(2)}$ with aperture opening angle is not merely a relative trend: Since the correlation function is determined directly rather than normalized to its asymptotic value, the observed suppression is shown to reach a clearly sub-Poissonian regime on an absolute scale.\\

Finally, we show that a combination of spatial and energetic filtering provides markedly reduced values of $g^{(2)}$ and also an improved trade-off regarding the according filter losses: Fig. 4(c), with analogous coloring to the other panels, presents the same spatially filtered dataset as (b), but imposing an 2\,eV energy filter centered at 1\,eV additionally. While in (b) the $g^{(2)}$ reduction saturates at a transmitted fraction of 10\%, the same fraction results in 4 times lower $g^{(2)}$ for the combined filter. Going to 3\% transmitted fraction, yields an unprecedented low $g^{(2)}=0.021$. For even stronger filtering no clear saturation of $g^{(2)}$ reduction can be observed before our $g^{(2)}$ estimate becomes shot noise dominated by the few number of multi-electron events (see logarithmic depiction in the inset). This trend seems again confirmed by the classical simulations: $g^{(2)}$ reaches below $10^{-4}$ before the onset of sampling noise.

\section*{Conclusion and outlook}
These results enable detailed predictions of the behavior of electrons emitted on ultrashort timescales from sharp needle sources, which are routinely employed in ultrafast electron microscopes. In such instruments, spatial filtering by apertures is unavoidable and leads to electron distributions that closely resemble our results for small angular acceptance. To date, electron microscopes have largely been designed with a focus on spatial resolution using continuous-wave electron beams. Our work provides, for the first time, a complete view of spatial and energy correlations simultaneously, that is, of transverse and longitudinal effects on equal footing. We expect these results to be instrumental for the design of dedicated electron columns optimized for ultrafast electron microscopy.

We have measured joint energy–angle correlations of two electrons emitted from metallic needle tips, constituting a three-dimensional study of their mutual Coulomb interaction. In the limiting cases, we observe a pronounced energy gap at small angular separations and strong angular repulsion for small energy differences, both of which are well reproduced by semiclassical simulations. Because the electrons act as ideal nanometric probes of one another, the resulting three-dimensional correlation signal encodes rich information about the spatiotemporal emission profile at the tip. While a full disentanglement of all contributing parameters from the correlation maps is challenging, we identify clear and systematic trends associated with specific experimental parameters through comparison with simulations. Moreover, by combining energy and spatial filtering, we demonstrate largely increased control over the electron number distributions of free-electron beams, both achieving unprecedented low $g^{(2)}$ values and providing a better trade-off of $g^{(2)}$ reduction and filter losses.

Looking ahead, Coulomb interactions may be exploited not only to shape emission statistics but also to enable emission-time selection of multi-electron events via energy filtering. For few-cycle laser pulses, our semiclassical simulations indicate that angular filtering can reduce the emission-time difference to the level of a few tens of attoseconds. Such experiments would provide stringent tests of existing theoretical models, both semiclassical and fully quantum mechanical. Ultimately, the resulting extreme phase-space degeneracies—corresponding to a high occupancy of available states-may even give rise to sub-Poissonian electron beams due to Pauli blocking, and even the two-electron state seen here might well be entangled.
\begin{acknowledgments}
This research was supported by the Gordon and Betty Moore Foundation (iQCE, 11473) and the DFG (Leibniz Prize).  J.H.\ and F.L.H.\ acknowledge support by the Max Planck School of Photonics.
\end{acknowledgments}
\appendix
\section{Detection losses and error bounds on $g^{(2)}$}
\label{sec:lossAppendix}
\begin{figure}
    \centering
    \includegraphics[width=\linewidth]{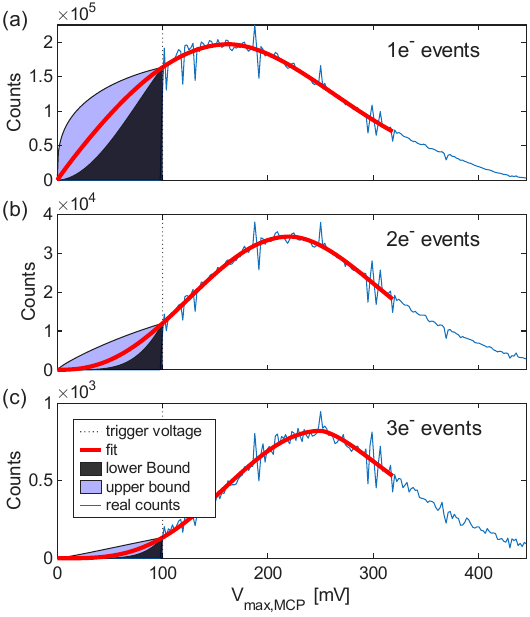}
    \caption{\fontsize{11}{12}\selectfont{\textbf{Trigger loss estimation fit by event electron number:} The maximum amplitude of the MCP's voltage trace after an electron emission event follows a broad distribution (blue lines), but depends on the electron number (compare panels). A fit (red line) helps to estimate the number of lost events due to an amplitude below the triggering voltage of 100\,mV. The shaded areas provide error bounds on this loss estimate}}
    \label{fig:mcpMaxVals}
\end{figure}

Acquisition of electron emission events is triggered only when the voltage pulse induced in the micro-channel-plate (MCP) exceeds a threshold voltage \cite{Knipfer2024}. Because the MCP gain is stochastic, the pulse amplitudes follow a broad distribution (blue lines in Fig. \ref{fig:mcpMaxVals}), and the fraction of events below the trigger threshold $V_\mathrm{thr}=100$\,mV stays undetected.\\ 
Losses do not bias observables that are uncorrelated with the detection probability, which is generally the case for measurements restricted to a fixed electron number $n$. However, the random pulse amplitude of an $n$-electron event is approximately given by the largest of $n$ random single-electron amplitudes (or more when pulses overlap), shifting the amplitude distribution to larger values with increasing $n$ (Fig. \ref{fig:mcpMaxVals}). Detection losses must therefore be accounted for when comparing event rates of different electron numbers $n$. In contrast, quantum-efficiency losses, i.e., the probability to initiate an electrode cascade, are assumed independent for different electrons and thus leave $g^{(2)}$ unchanged \cite{MandelWolf1995}.\\
Detection losses are estimated by fitting the $V_\mathrm{max}$ histograms for each $n$ with a quarter sine (left) combined with half of a Gaussian (right),
$$f(V_\mathrm{max};\mu,\sigma,A,B,C)=A\begin{cases}\sin(\frac{\pi V_\mathrm{max}}{2\mu})^B,~~V_\mathrm{max}<\mu\\\exp(\frac{(V_\mathrm{max}-\mu)^C}{\sigma}),~~V_\mathrm{max}\geq\mu\end{cases},$$
both with a distorted curvature via a fitted exponent (see thick red lines in Fig. \ref{fig:mcpMaxVals}). Integrating the fits below $V_\mathrm{thr}$ yields our loss estimates. They are listed in Table \ref{tab:losses} together with error bounds obtained by a nonlinear distortion of the fit curve below $V_\mathrm{thr}$ (shaded areas in Fig. \ref{fig:mcpMaxVals}).\\
$g^{(2)}$ is calculated from energy- or angle-filtered rates of events with $n=1,2,3$ electrons; rare higher-number events are negligible at the present level of precision. We note that partially filtered $n$-electron events still require a detection efficiency correction according to their unfiltered number. The error boundaries of $g^{(2)}$ are obtained by maximizing and minimizing it over the entire 3D space of possible loss values for $n=1,2,3$.\\
For the measurement shown in Fig. \ref{fig:figure4} (b), the average post-selected rate reaches up to 0.2 electrons per pulse. At such rates, a narrow error bound on $g^{(2)}$ could only be obtained by reducing $V_\mathrm{thr}$ to 20\,mV, lowering losses to 3.5\%, 0.13\% and 0.005\% for $n=1,2,3,$ respectively. At the same time triple-electron events become non-negligible. Since the position/energy reconstruction of such events is less bench-marked than for $n=1,2,$ a further conservative boundary is adopted: the maximum fraction of triples within each aperture angle is derived from the one expected for 3 uncorrelated single electrons, i.e., we assume that the beam of $n=3$ events is at least as divergent as the beam of $n=1$ events.
\newline
\begin{table}[]
    \centering
    \begin{tabular}{|l|c|c|c|}
        \hline
        $n$ & 1 & 2 & 3\\ 
        \hline
        Losses [\%] & 19.2 & 5.0 & 1.9 \\
        \hline
        Lower bound [\%]& 14.4 & 3.0 & 1.1 \\
        \hline
        Upper bound [\%]& 25.1 & 8.7 & 3.8 \\
        \hline
    \end{tabular}
    \caption{\fontsize{11}{12}\selectfont{\textbf{Trigger loss estimate values:} The trigger losses for events of $n=1,2,3$ electrons and an error bound for them are extracted by the fits in Fig. \ref{fig:mcpMaxVals}.}}
    \label{tab:losses}
\end{table}

\section{Measurement at closer detector distance}
\label{sec:meas2Appendix}
\begin{figure}[H]
    \centering
    \includegraphics[width=\linewidth]{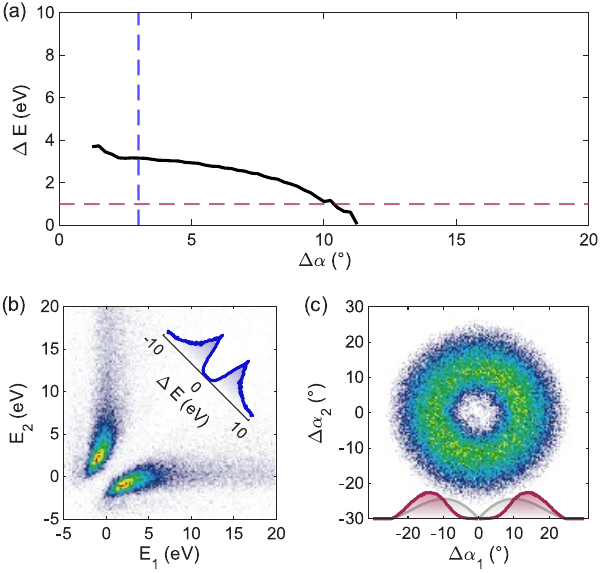}
    \caption{\fontsize{11}{12}\selectfont{\textbf{AEDS and filtered distribution of second measurement:} Analogous to Fig. \ref{fig:figure2} (a)-(c), with reduced detector distance and a new sample.}}
    \label{fig:meas2}
\end{figure}

A second measurement at a closer detector distance (177\,mm instead of 247\,mm) allowed for a higher range of absolute emission angles, for a better characterization of $g^{(2)}$ over aperture opening in Fig. \ref{fig:figure4} (b). Due to this and other modification work is the experimental vacuum chamber, a new sample of presumably slightly higher radius, yielding a lower field enhancement, has to be used. This may explain the differences in the AEDS and ERL measured for this tip, see Fig. \ref{fig:meas2}.

\bibliography{literature}

\end{document}